\documentclass[twocolumn,amsmath,amssymb,prl]{revtex4}

\usepackage{graphicx}
\usepackage{dcolumn}
\usepackage{bm}
\usepackage{color}
\usepackage{notes2bib}

\begin{document}

\title{Influence of irradiation on defect spin coherence in silicon carbide}

\author{C.~Kasper$^{1}$}
\author{D.~Klenkert$^{1}$}
\author{Z.~Shang$^{2}$}
\author{D.~Simin$^{1}$}
\author{A.~Sperlich$^{1}$}
\author{H.~Kraus$^{3}$} 
\author{C.~Schneider$^{4}$}
\author{S.~Zhou$^{2}$}
\author{M.~Trupke$^{5}$}
\author{W.~Kada$^{6}$}
\author{T.~Ohshima$^{7}$}
\author{V.~Dyakonov$^{1}$}
\author{G.~V.~Astakhov$^{2,1}$}
\email[E-mail:~]{g.astakhov@hzdr.de}

\affiliation{$^1$Experimental Physics VI, Julius-Maximilian University of W\"{u}rzburg, 97074 W\"{u}rzburg, Germany \\
$^2$Helmholtz-Zentrum Dresden-Rossendorf, Institute of Ion Beam Physics and Materials Research, 01328 Dresden, Germany  \\
$^3$Jet Propulsion Laboratory, California Institute of Technology, Pasadena, CA 91109, USA \\
$^4$Helmholtz-Zentrum Dresden-Rossendorf, Institute of Radiation Physics, 01328 Dresden, Germany \\
$^5$Vienna Center for Quantum Science and Technology, Universit\"{a}t Wien, 1090 Vienna, Austria \\
$^6$Faculty of Science and Technology, Gunma University, Kiryu, Gunma 376-8515, Japan \\
$^7$National Institutes for Quantum and Radiological Science and Technology, Takasaki, Gunma 370-1292, Japan }

\begin{abstract}
Irradiation-induced lattice defects in silicon carbide (SiC) have already exceeded their previous reputation as purely performance-inhibiting. With their remarkable quantum properties, such as long room-temperature spin coherence and the possibility of downscaling
to single-photon source level, they have proven to
be promising candidates for a multitude of quantum information applications. One of the most crucial parameters of any quantum system is how long its quantum coherence can be preserved. By using the pulsed optically detected magnetic resonance (ODMR) technique, we investigate the spin-lattice relaxation time ($T_1$) and spin coherence time ($T_2$) of silicon vacancies in 4H-SiC
created by neutron, electron and proton irradiation in a broad range
of fluences. We also examine the effect of irradiation energy and
sample annealing. We establish a robustness of the $T_1$ time against all types of irradiation and reveal a universal scaling of the $T_2$ time with the emitter density. Our results can be used to optimize the coherence properties of silicon vacancy qubits in SiC  for specific tasks.
\end{abstract}

\date{\today}

\maketitle

\section*{Introduction} 

Defect spins in silicon carbide (SiC) attract growing interest because of their high potential as multidimensional qubits -- the so called qudits -- for wafer-scale quantum information processing at room temperature \cite{Baranov:2011ib, Koehl:2011fv, Riedel:2012jq, Weber:2010cn}. Particularly, silicon vacancies ($\mathrm{V_{Si}}$) and divacancies ($\mathrm{VV}$) demonstrate an extremely long spin-lattice relaxation time ($T_1$) and spin coherence time ($T_2$), even in commercial wafers without isotope purification  \cite{Falk:2013jq, Yang:2014kqa, Carter:2015vc, Seo:2016ey, Simin:2017iw, Fischer:2018fj, Brereton:2018ur, Soltamov:2019hr}. These defects can be used for room-temperature quantum metrology \cite{Kraus:2013di, Falk:2014fh, Soykal:2016tk, Castelletto:2013jj}, including magnetometry \cite{Kraus:2013vf, Simin:2015dn, Lee:2015ve, Simin:2016cp, Niethammer:2016bc, Cochrane:2016dd} and thermometry \cite{Kraus:2013vf, Anisimov:2016er}. Furthermore, single defects have been isolated, which can be used as single photon emitters \cite{Castelletto:2013el, Fuchs:2015ii}, and coherent control of single defect spins  \cite{Christle:2014ti, Widmann:2014ve} with high fidelity \cite{Christle:2017tq, Nagy:2019fw, Banks:2018ve} has been realized.  

SiC itself is a technologically highly developed wide band gap semiconductor with unique mechanical, electrical and optical properties, that make it very attractive for various electronic and optoelectronic applications under extreme conditions. Therefore, it is a unique platform to implement hybrid quantum systems, where defect spins can be coupled to the eigenmodes of mechanical resonators \cite{Whiteley:2019eu, Poshakinskiy:2019wd} and photonic cavities \cite{Calusine:2014gv, Bracher:2015gg, Radulaski:2017ic, Lohrmann:2017ks} or integrated into photoelectronic circuits \cite{Fuchs:2013dz, Lohrmann:2015hd, Niethammer:2019uq}.  Because SiC is a bio-compatible material, the fabrication of spin-carrying defects in SiC nanocrystals \cite{Castelletto:2014eu, Muzha:2014th} gives an opportunity for \textit{in-vivo} imaging of chemical processes. 

\begin{figure}[t]
\includegraphics[width=.48\textwidth]{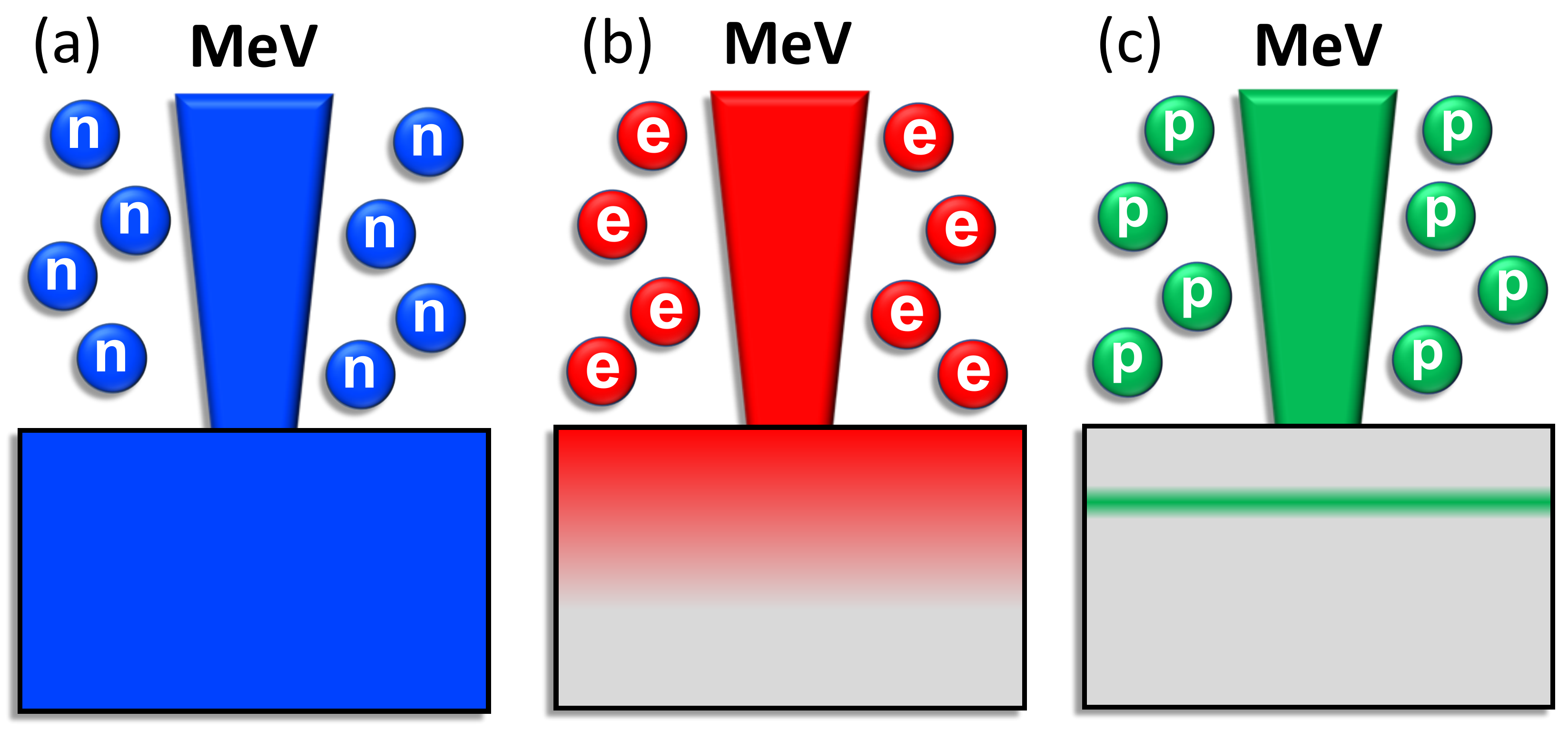}
\caption{Schematic generation of the $\mathrm{V_{Si}}$ defects in SiC by MeV energy particle irradiation. (a) Neutron irradiation leads to a homogeneous distribution of the $\mathrm{V_{Si}}$ defects in the crystal. (b) Electrons show a penetration depth depending on the electron energy. In case of thick samples, this may lead to a gradient of the $\mathrm{V_{Si}}$ density with highest density close to the surface.  (c) Ion irradiation (e. g. protons) results in the formation of a $\mathrm{V_{Si}}$ layer at a depth depending on the ion energy. } \label{fig1}
\end{figure}

Various approaches can be used to create intrinsic defects in SiC, including thermal quenching \cite{Baranov:2011ib} and laser writing  \cite{Chen:2018vp}. However, the most common and frequently used methods are based on electron \cite{Sorman:2000ij, vonBardeleben:2000jg}, neutron \cite{Heinisch:2004ku, Fuchs:2015ii} and ion \cite{vonBardeleben:2000es, Kraus:2017cka, Wang:2017fb, Wang:2016vn, Embley:2017bf, Wang:2018uo} irradiation.  There is no optimum irradiation method, as they all have their own advantages and disadvantages, which are schematically presented in Fig~\ref{fig1}. 

The irradiation with neutrons generates a homogeneous distribution of silicon vacancies in bulk crystals (Fig.~\ref{fig1}a). In combination with high irradiation fluences, this method allows to create high-density $\mathrm{V_{Si}}$ ensembles, as required for the realization of room-temperature masers \cite{Kraus:2013di} and highly sensitive magnetic field sensors \cite{Simin:2016cp}. Since neutrons are not charged, they lose almost all of their energy in collisions with nuclei in the lattice and hence considerably damage the SiC crystal. This leads to the creation of a multitude of defect types other than $\mathrm{V_{Si}}$ up to the formation of defect clusters, resulting in a deterioration of the $\mathrm{V_{Si}}$ coherence properties.

A more gentle approach is electron irradiation (Fig.~\ref{fig1}b) as a large part of their energy is lost in interactions with crystal electrons. However, the electron damage cannot (in reasonable beam time with common accelerators) reach the quantum center densities achievable after irradiation with neutrons in nuclear reactors. Furthermore, the electron irradiation leads to a quantum center density gradient in the sample, with highest density close to the surface. The electron penetration depth depends on its kinetic energy. For thin samples (up to  $100 \, \mathrm{\mu m}$) and $\mathrm{MeV}$-electron energy, the inhomogeneity of the in-depth defect distribution is marginal and can be neglected \cite{Campbell:2000hl}. However for thick ($> 1 \, \mathrm{mm}$) samples, the in-depth inhomogeneous distribution should be taken into account. The samples in this study have a thickness of about $300 \, \mathrm{\mu m}$.

In contrast to the aforementioned methods, ions show an increasing stopping power for lower energies according to the Bethe-Bloch equation \cite{Sigmund:2006fd}.  This results in the formation of a defect layer at a depth corresponding to the so-called Bragg peak (Fig.~\ref{fig1}c). This approach can be used to write 3D defect structures in a SiC crystal using a focussed ion beam \cite{Kraus:2017cka}, allowing the creation of defect spins in the desired place of a nanostructure. On the other hand, the ion irradiation is not suitable for the creation of defect spin ensembles with an in-depth homogeneous distribution. 

Any type of irradiation results in the creation of various types of defects, which also change the coherence properties of the vacancy-related spins. The impact of the irradiation on the $T_1$ and $T_2$ times, in spite of their crucial importance for quantum applications, has not been systematically investigated so far. In this work, we compare the $\mathrm{V_{Si}}$ spin relaxation and spin coherence in SiC created by neutron, electron and proton irradiation. Particularly, $T_1$ and  $T_2$ times are measured in a broad range of the density of optically active quantum centers in the infrared regime (from here on referred to as emitter density $N_{\mathrm{V}}$), from dense ensembles down to isolated defects. To  improve the coherence of the $\mathrm{V_{Si}}$ defects after irradiation, we also examine the influence of sample annealing. Our results allow to predict the $\mathrm{V_{Si}}$ spin coherence properties depending on the irradiation type applied, fluence and particle energy as well as to optimize the irradiation parameters for a variety of quantum applications.

\section*{Results} 
  
\begin{figure}[t]
\includegraphics[width=.48\textwidth]{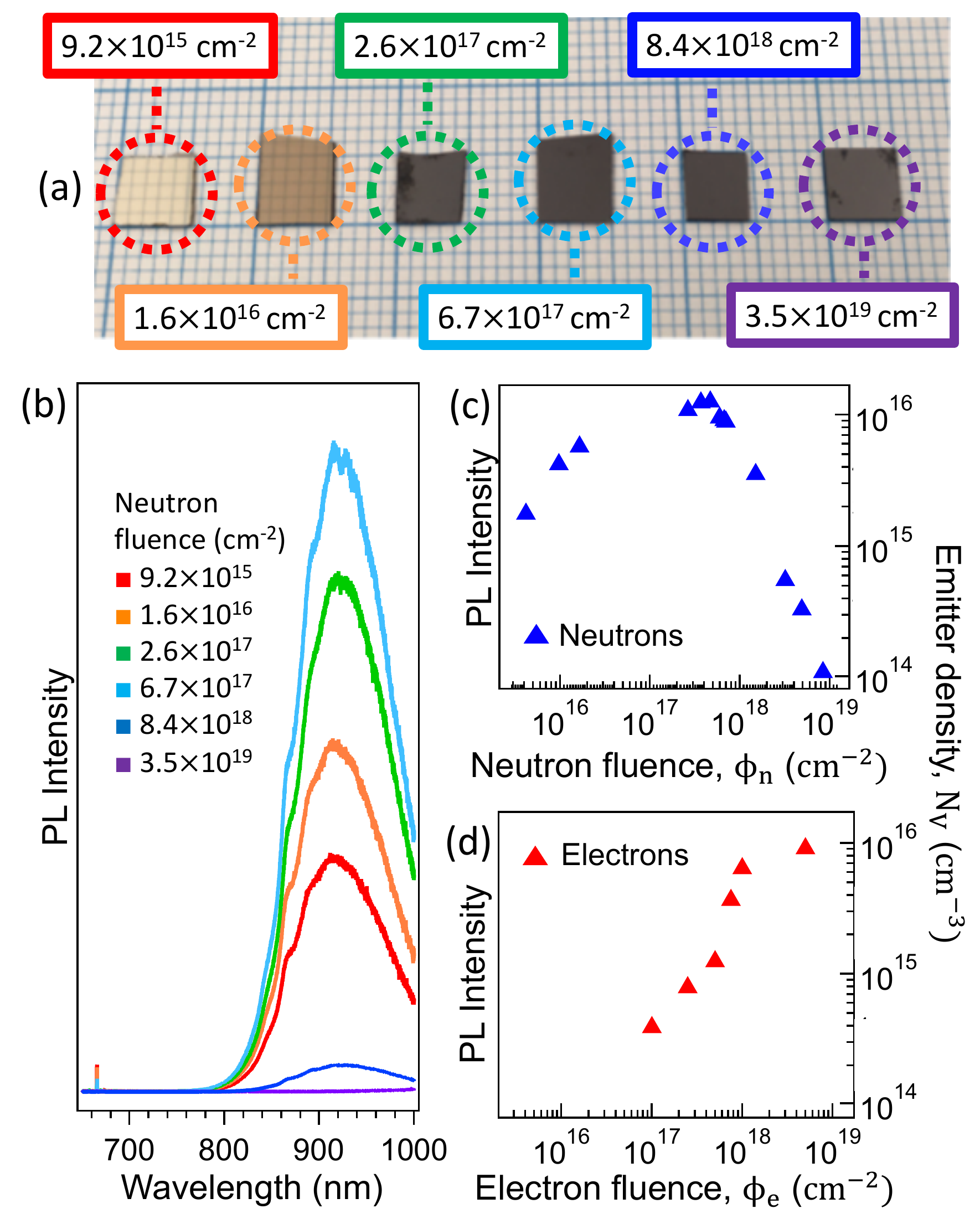}
\caption{Irradiation impact on the optical properties of SiC. (a) Photographs of the SiC crystals for different neutron irradiation fluences. (b) PL spectra of the $\mathrm{V_{Si}}$ defects for different neutron irradiation fluences. (c) Variation of the PL intensity and corresponding emitter density as a function of the irradiation fluence with epithermal and fast neutrons. (d) Same as (c) for $2 \, \mathrm{MeV}$ electrons.} \label{fig2}
\end{figure}  

Figure~\ref{fig2}a shows the modification of the optical properties of a SiC crystal with increasing neutron irradiation fluence. The irradiation was performed in nuclear research reactors. Since thermal neutrons only produce negligible displacement, only epithermal and fast (MeV) neutrons are counted in the fluence calculation \cite{Wendler:2012co}. In our experiments, we use commercial high-purity semi-insulating (HPSI) 4H-SiC wafers.  SiC is a wide bandgap semiconductor with a bandgap of $3.23 \, \mathrm{eV}$ and is therefore transparent in case of the lowest neutron fluence $\Phi_n = 9.2 \times 10^{15} \, \mathrm{cm^{-2}}$. Neutron irradiation creates a variety of optically active and inactive defects, which has a direct effect on the optical properties of the crystal due to absorption and scattering of light by the defects. High irradiation fluences cause crystal destruction and in case of the highest neutron fluence $\Phi_n = 3.5 \times 10^{19} \, \mathrm{cm^{-2}}$, SiC becomes completely opaque. The generation of the $\mathrm{V_{Si}}$ defects is verified by their characteristic room-temperature photoluminescence (PL) band in the near infrared (NIR) spectral range of $800 - 1000 \, \mathrm{nm}$  \cite{Hain:2014tl}, as shown in  Fig.~\ref{fig2}b. The $\mathrm{V_{Si}}$ PL intensity  shows non-monotonic behavior with the irradiation fluence. This is because high irradiation fluences lead to crystal damage and the creation of other types of defects (including defect clusters) providing non-luminescent relaxation paths for the $\mathrm{V_{Si}}$ defects. 

To find the emitter density $N_{\mathrm{V}}$, we calibrate the PL intensity in our confocal setup with a defined collection volume using a 4H-SiC sample with known $N_{\mathrm{V}}$  \cite{Fuchs:2015ii}.  Figure~\ref{fig2}c presents the concentration of the  optically active $\mathrm{V_{Si}}$ defects as a function of the neutron fluence, which is varied over more than three orders of magnitude. The maximum $N_{\mathrm{V}}  = 2 \times 10^{16}  \, \mathrm{cm^{-3}}$ is observed for  $\Phi_n =  6.7 \times 10^{17} \, \mathrm{cm^{-2}}$ and dramatically drops down for the higher neutron fluences. The electron irradiated samples reveal qualitatively similar behavior. Figure~\ref{fig2}d presents $N_{\mathrm{V}}$ as a function of the electron fluence $\Phi_e$ for an electron energy of $E_e = 2 \, \mathrm{MeV}$. The maximum $N_{\mathrm{V}}  = 1 \times 10^{16}  \, \mathrm{cm^{-3}}$ is achieved for  $\Phi_e =  4 \times 10^{18} \, \mathrm{cm^{-2}}$. 
A higher $\Phi_e$ than $\Phi_n$  is required to achieve the respective maximum $N_{\mathrm{V}}$ in our samples, pointing at a lower $\mathrm{V_{Si}}$ creation yield in case of electron irradiation compared to neutron irradiation. 

\begin{figure}[t]
\includegraphics[width=.44\textwidth]{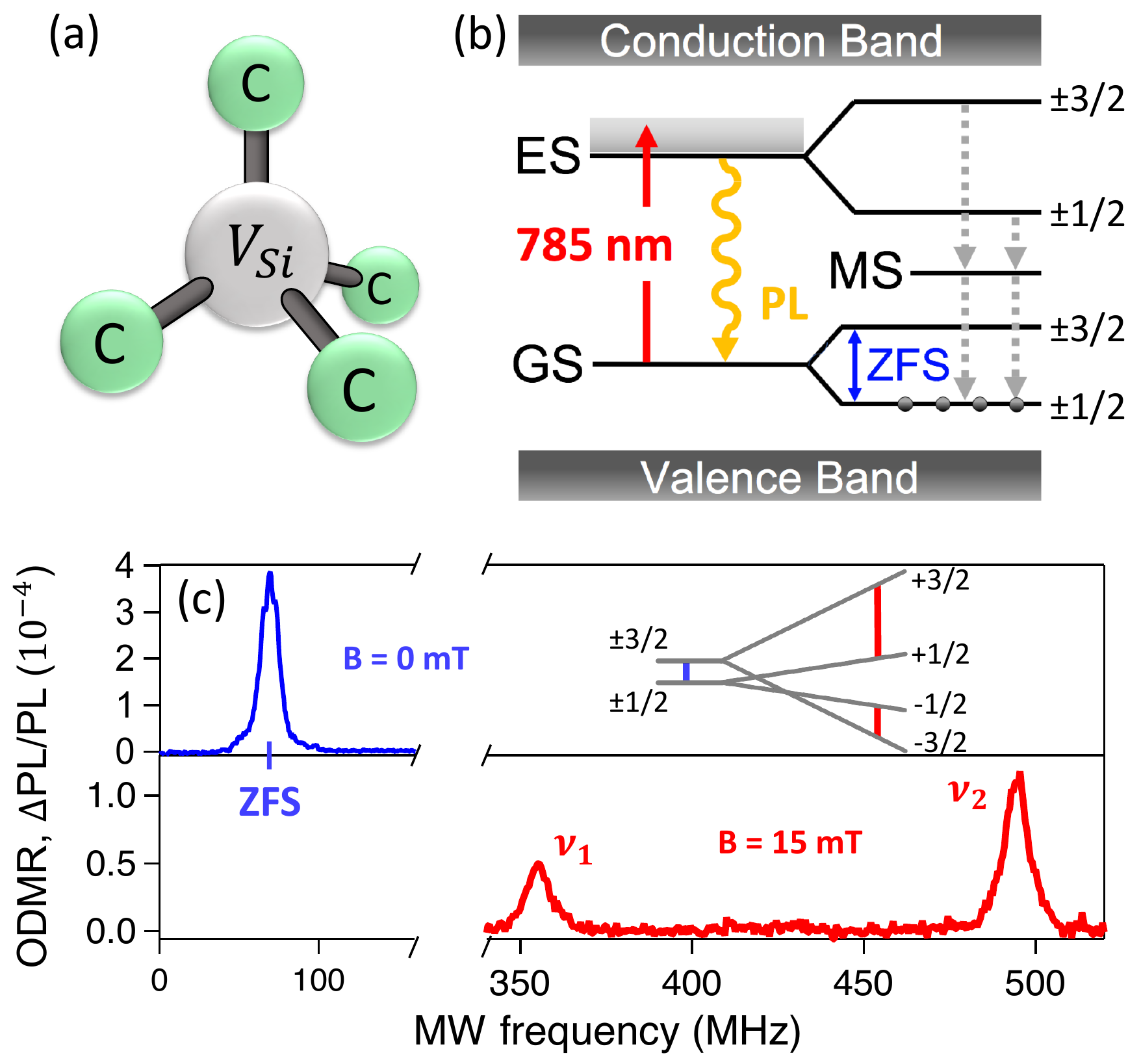}
\caption{(a) Schematic presentation of a single $\mathrm{V_{Si}}$ defect. (b) Energy and spin structure of the negatively charged $\mathrm{V_{Si}}$ defect with  $S = 3/2$.  The solid arrow represents laser excitation, the wave arrow represents the radiative recombination and the dashed arrows represent the spin-dependent relaxation. (c) Room temperature ODMR spectra of a $\mathrm{V_{Si}}$ ensemble without applied magnetic field (blue) and in an external magnetic field $B = 15 \, \mathrm{mT}$ (red). The inset shows the Zeeman splitting of the spin sublevels with corresponding transitions represented by the vertical lines.} \label{fig3}
\end{figure}

The crystallographic $\mathrm{V_{Si}}$ configuration is schematically presented in Fig.~\ref{fig3}a. The capture of an electron by a silicon vacancy, in addition to the four electrons from the surrounding carbon atoms  leads to a negatively charged state. Only this charge configuration of $\mathrm{V_{Si}}$ is optically active in NIR.  Two of the total five electrons build a spin singlet and the three remaining electrons form a spin quadruplet with $S = 3/2$ \cite{Kraus:2013di}. Upon laser excitation at $785 \, \mathrm{nm}$, the radiative recombination from the excited state (ES) to the ground state (GS) leads to the characteristic NIR PL of the $\mathrm{V_{Si}}$ defects, as presented in the energetic structure of Fig.~\ref{fig3}b. There are two types of the $\mathrm{V_{Si}}$ defect in the 4H-SiC polytype, referred to as V1 and V2 centers \cite{Sorman:2000ij}. 

In this study, we concentrate on the V2 center with the characteristic zero-phonon line (ZFL) at $ 917 \, \mathrm{nm}$ and the zero-field splitting (ZFS) of $70 \, \mathrm{MHz}$ between the GS spin sublevels $m_S= \pm 1/2$ and $m_S = \pm 3/2$. There is also spin-dependent recombination from the ES to the GS through a metastable state (MS), resulting in the preferential population of the $m_S= \pm 1/2$ spin sublevel (Fig.~\ref{fig3}b). The PL intensity is higher for the $m_S= \pm 3/2$ spin state, which is the basis for the optically detected magnetic resonance (ODMR) experiments with microwave (MW) induced transitions between the GS spin sublevels \cite{Kraus:2013vf}. In the absence of an external magnetic field ($B = 0 \, \mathrm{mT}$), the ZFS can directly be observed in the ODMR spectrum  (Fig.~\ref{fig3}c).  The ODMR contrast is calculated as the MW induced PL change ($\mathrm{\Delta PL}$) divided by the PL intensity, i.e., $\mathrm{\Delta PL / PL}$.   In an external magnetic field ($B \neq 0 \, \mathrm{mT}$), the GS spin sublevels are split as shown in the inset of Fig.~\ref{fig3}c. This results in two ODMR lines at the MW frequencies $\nu_{1,2} =  | \mathrm{ZFS} \pm \gamma B |$, where $\gamma = 28 \, \mathrm{MHz / mT}$ is the electron gyromagnetic ratio. An example of an ODMR spectrum at $B = 15 \, \mathrm{mT}$ is shown in Fig.~\ref{fig3}c. 

\begin{figure}[t]
\includegraphics[width=.48\textwidth]{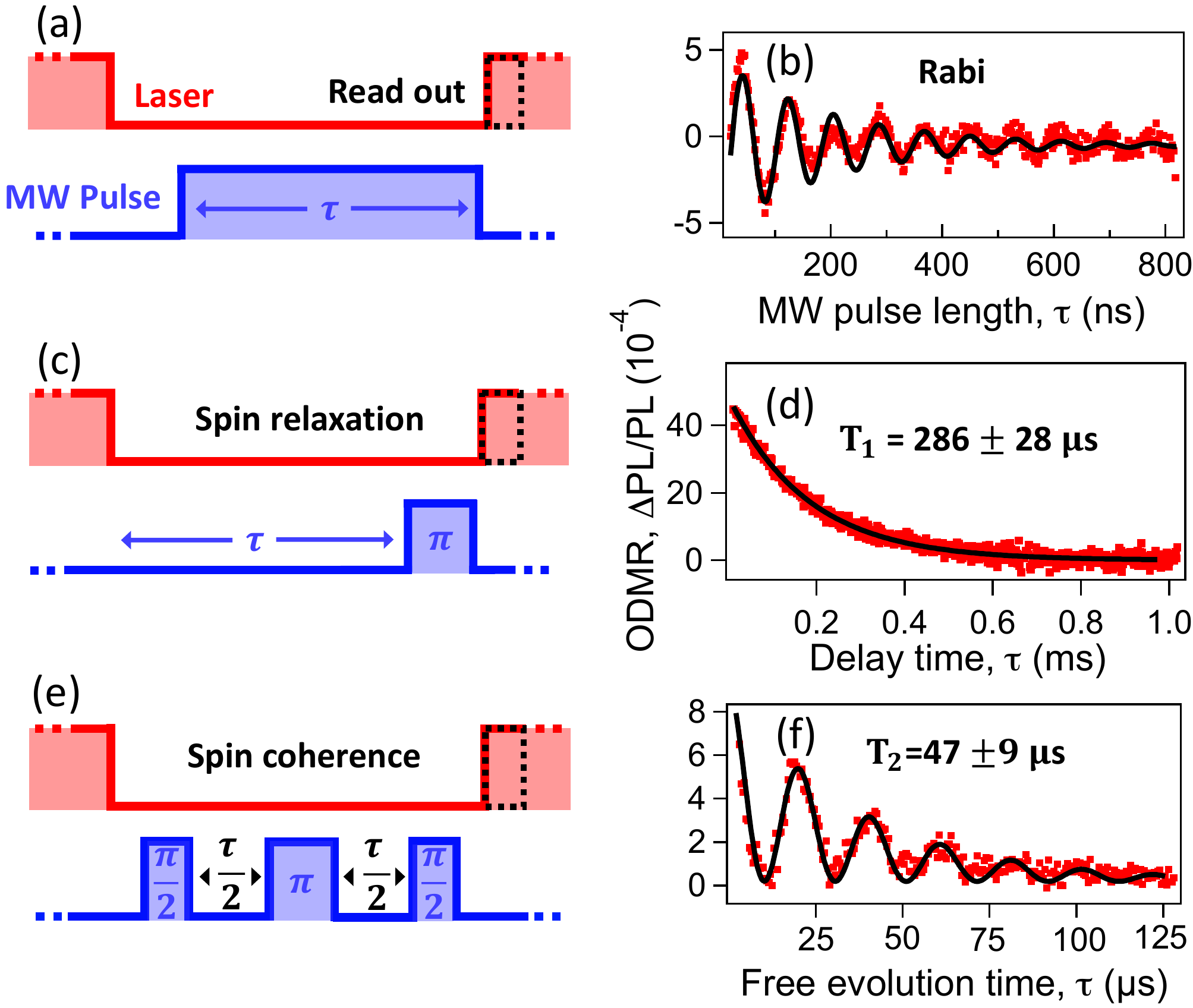}
\caption{Coherent control of the $\mathrm{V_{Si}}$ defects. (a), (c), (e) Laser and MW pulse sequences  to measure Rabi oscillations, spin lattice relaxation time $\mathrm{T_1}$ and spin echo coherence time $\mathrm{T_2}$, respectively. For read out of the spin state we use a second laser pulse and collect the PL (dashed box), where $\tau$ is a variable parameter. (b), (d), (f) The symbols represent experimental data for Rabi oscillations as well as for T1 and T2, respectively.  The solid lines are fitting curves as discussed in the main text. The data are obtained at room temperature in a magnetic field of $15 \, \mathrm{mT}$. The measured 4H-SiC sample is irradiated with electrons to a fluence  $\Phi_e =  1 \times 10^{17} \, \mathrm{cm^{-2}}$ corresponding to  $N_{\mathrm{V}}  = 3.9 \times 10^{14}  \, \mathrm{cm^{-3}}$.} \label{fig4}
\end{figure}

To measure the coherence properties of the $\mathrm{V_{Si}}$ spins at room temperature, we use pulsed ODMR as schematically presented in Figs.~\ref{fig4}a, c and e. The pulse sequences are described in detail in earlier works \cite{Simin:2017iw, Soltamov:2019hr}. The first laser pulse is used to initialize the system into the $m_S= \pm 1/2$ state. It is followed by a MW pulse to drive the $\nu_2$ spin transition between the $+1/2$ and $+3/2$ spin sublevels (inset of Fig.~\ref{fig3}c). We apply a magnetic field $B = 15 \, \mathrm{mT}$ along the $c$-axis of the 4H-SiC crystal to suppress the heteronuclear-spin flip-flop processes and hence the decoherence \cite{Yang:2014kqa, Seo:2016ey}. The second laser pulse is used to reinitialize the system and read out the spin states of the $\mathrm{V_{Si}}$ defects.  For this purpose, $\mathrm{\Delta PL/PL}$, which corresponds to the spin states of the $\mathrm{V_{Si}}$ defects, is detected immediately after the laser is switched on (dashed box in  Figs.~\ref{fig4}a, b and d).  We first measure Rabi oscillations by varying the MW pulse length $\tau$ (Fig.~\ref{fig4}a). An example of a measurement on an electron irradiated sample with $\Phi_e =  1 \times 10^{17} \, \mathrm{cm^{-2}}$ corresponding to  $N_{\mathrm{V}}  = 3.9 \times 10^{14}  \, \mathrm{cm^{-3}}$ is presented in Fig.~\ref{fig4}b. The fitting of these data to an exponentially decaying sine wave (solid line) allows to calibrate the MW duration for the $\pi$- and $\pi/2$-pulses.  

To measure the spin-lattice relaxation time $T_1$ we use the protocol of Fig.~\ref{fig4}c, where the MW $\pi$-pulse is applied after the varyable delay time $\tau$ to exchange the spin population between the $m_S = + 3/2$ and $m_S = +1/2$ states. The fitting of the experimental data to a mono-exponential decay of $\exp (- 2 \tau / T_1)$ gives the spin relaxation time $T_1$ \cite{Widmann:2014ve}.  A fitting example is shown in Fig.~\ref{fig4}d by the solid line. We note that the spin relaxation time between the $m_S = + 1/2$ and $m_S = -1/2$ states is expected to be shorter $T^{\prime}_1 = 3 T_1 / 4$ \cite{Soltamov:2019hr}.  

To measure the spin coherence time $T_2$, we use the spin echo sequence presented in  Fig.~\ref{fig4}e. The first $\pi/2$-pulse creates a coherent superposition of the $m_S = + 3/2$ and $m_S = +1/2$ states. It is followed by a $\pi$-pulse to refocus the spin coherence after dephasing and the last $\pi/2$-pulse to project the created superposition to an optically readable state. An example of the experimental data is presented in Fig.~\ref{fig4}f. In addition to an exponential decay, the signal is modulated with a few frequency components, known as the electron spin-echo envelope modulation (ESEEM). These modulations  have been shown to arise through hyperfine interactions between the $\mathrm{V_{Si}}$ defect spins and the surrounding $\mathrm{^{29}Si}$ and $\mathrm{^{13}C}$ nuclei \cite{Widmann:2014ve, Christle:2014ti, Carter:2015vc}.  In case of $S = 3/2$, the modulation frequencies and modulation amplitudes reveal a complex magnetic field dependence.  To fit the experimental data, we use the product of oscillating functions decaying as $\exp (- \tau / T_2)$. The details of the fitting procedure are described elsewhere \cite{Simin:2017iw}.  An example of such a fit (solid line), from which we determine $T_2$ is shown in  Fig.~\ref{fig4}f.

\section*{Discussion} 

\subsection*{Effect of irradiation} 

\begin{figure}[t]
\includegraphics[width=.48\textwidth]{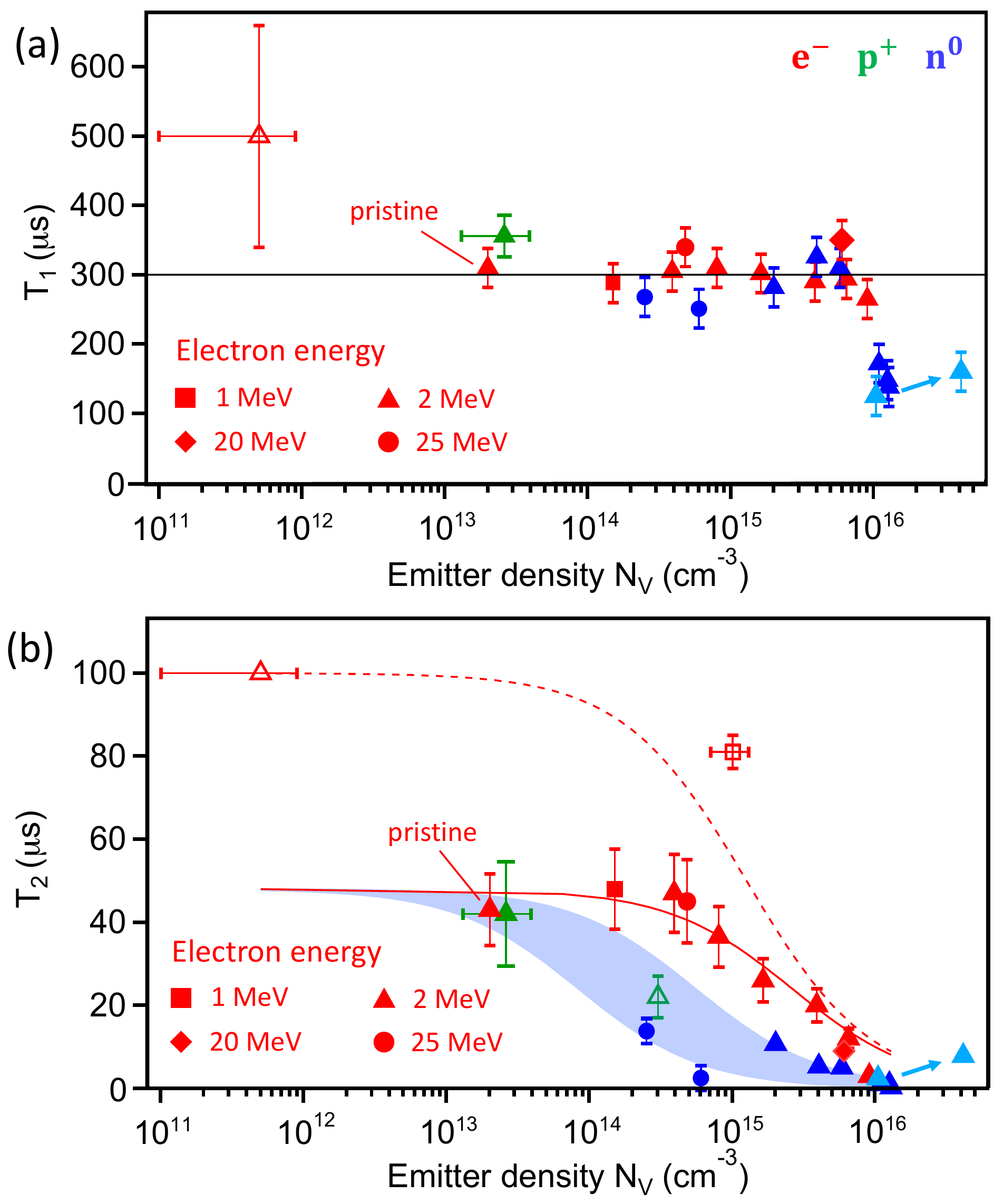}
\caption{ Spin coherence properties of the $\mathrm{V_{Si}}$ defects in 4H-SiC for different irradiation particles  as a function of  emitter density. The samples were irradiated with electrons, neutrons or protons.  (a) Spin-lattice relaxation time $\mathrm{T_1}$ as a function of the emitter density. The red close symbols correspond to our measurements with electron energies $1 \, \mathrm{MeV}$ (squares),  $2 \, \mathrm{MeV}$ (triangles), $25 \, \mathrm{MeV}$ (circles) and $20 \, \mathrm{MeV}$ (diamonds). The green triangle corresponds to proton irradiation with an energy of $1.7 \, \mathrm{MeV}$ \cite{Kraus:2017cka}. The red open triangle corresponds to a single $\mathrm{V_{Si}}$ defect \cite{Widmann:2014ve}. The $\mathrm{T_1}$ time is a constant around  $300 \, \mathrm{\mu s}$ for low emitter densities (the solid line) and drops rapidly for high densities due to crystal damage. The blue closed triangles and circles correspond to the neutron irradiation in the BER II reactor in Berlin and in a TRIGA Mark-II reactor in Vienna, respectively. The cyan triangles represent measurements on a neutron irradiated sample before and after annealing (Fig. 7 a). (b) The spin coherence time $\mathrm{T_2}$ as a function of emitter density. The assignment of symbols is the same as in (a).  The red (green) open square (triangle) corresponds to an electron (proton) irradiation energy of  $2 \, \mathrm{MeV}$ \cite{Carter:2015vc, Embley:2017bf}. The red solid and dashed lines as well as the cyan area are fits to Eq.~(\ref{T2_Phi}) with parameters described in the text. } \label{fig5}
\end{figure}

Figure~\ref{fig5}a summarizes the measurements of the spin-lattice relaxation time $T_1$ as a function of the emitter density ($N_{\mathrm{V}}$) in electron, neutron and proton irradiated samples. It is quite robust against irradiation for all irradiation types up to $N_{\mathrm{V}} = 7 \times 10^{15}  \, \mathrm{cm^{-3}}$ and equals $T_1 \sim 300 \, \mathrm{\mu s}$. Remarkably, the electron energy in the range from $1 \, \mathrm{MeV}$ to $25 \, \mathrm{MeV}$  has no effect on $T_1$.  The $\mathrm{V_{Si}}$ spin relaxation is dominated by the Raman mechanism of spin-phonon scattering \cite{Simin:2017iw, Poshakinskiy:2019wd}, which is an intrinsic property of a SiC crystal and hence independent of the irradiation fluence. This result corresponds well to $T_1 = 500 \pm 160 \, \mathrm{\mu s}$ reported for a single $\mathrm{V_{Si}}$ defect in a sample with  $N_{\mathrm{V}} \sim 10^{11}  - 10^{12}  \, \mathrm{cm^{-3}}$ \cite{Widmann:2014ve}. In case of higher emitter density, the spin relaxation is given by the interaction with surrounding defects and the $T_1$ time decreases rapidly with $N_{\mathrm{V}}$. 

The spin coherence time $T_2$ reveals strong dependence on the irradiation fluence as shown in Fig.~\ref{fig5}b. The spin decoherence  is governed by interactions with the surrounding spin bath of nuclei ($\mathrm{^{29}Si}$, $\mathrm{^{13}C}$) $1/T_2^{(n)}$, paramagnetic centers (particularly $\mathrm{^{14}N}$) $1/T_2^{(p)}$ as well as residual (particularly carbon vacancies) $1/T_2^{(r)}$ and irradiation-induced defects $1/T_2^{(i)}$ \cite{Yang:2014kqa, Carter:2015vc, Stanwix:2010ko}.  The resulting spin decoherence rate is then $1/ T_2 = 1/T_2^{(n)} + 1/T_2^{(p)} + 1/T_2^{(r)} + 1/T_2^{(i)}$. By applying a moderate magnetic field $1/T_2^{(n)}$ is suppressed  \cite{Yang:2014kqa}. We assume that $1/T_2^{(i)}$ is proportional to the number of all irradiation induced defects, which in turn scales with $N_{\mathrm{V}}$. Under this assumption, the effect of irradiation can be described by 
\begin{equation}
T_2 = \frac{T_2^{\mathrm{(pristine)}}}{1 + T_2^{\mathrm{(pristine)}} \kappa N_{\mathrm{V}}} \,.
\label{T2_Phi}
\end{equation} 
Here, $T_2^{\mathrm{(pristine)}}$ is the spin coherence time in a pristine, non-irradiated sample given by the presence of residual paramagnetic centers and nuclear spins. It is therefore depending on the intrinsic properties and quality of a wafer. The coefficient $\kappa$ describes the assumed proportionality $1/T_2^{(i)} = \kappa N_{\mathrm{V}}$. 

We now discuss the experimental data for $T_2$ in the electron irradiated samples presented in Fig.~\ref{fig5}b. As in case of the $T_1$ time, we do not find any significant dependence on the electron energy. In our HPSI wafers (solid symbols) with low irradiation fluence ($\Phi_e < 3 \times 10^{17} \, \mathrm{cm^{-2}}$, $N_{\mathrm{V}} = 7 \times 10^{14}  \, \mathrm{cm^{-3}}$), the $T_2$ time is nearly independent of $N_{\mathrm{V}}$.  The experimental data can be well fitted to Eq.~(\ref{T2_Phi}) with $T_2^{\mathrm{(pristine)}} = 48 \, \mathrm{\mu s}$ and $\kappa = 0.8 \times 10^{-11}  \, \mathrm{s^{ -1} cm^{3}}$ (solid line in  Fig.~\ref{fig5}b). A deviation towards lower values is observed for $N_{\mathrm{V}} > 7 \times 10^{15}  \, \mathrm{cm^{-3}}$, which correlates with the $T_1$ shortening in Fig.~\ref{fig5}a.

The $\mathrm{V_{Si}}$ spin coherence in some other SiC samples (open symbols in Fig.~\ref{fig5}b) was reported to be longer. It is $T_2  \sim 100 \,  \mathrm{\mu s}$ for a single $\mathrm{V_{Si}}$ with $N_{\mathrm{V}} \sim 10^{11}  - 10^{12}  \, \mathrm{cm^{-3}}$ in an epitaxial 4H-SiC layer \cite{Widmann:2014ve}.  Another HPSI 4H-SiC wafer after electron irradiation $\Phi_e = 5 \times 10^{17} \, \mathrm{cm^{-2}}$ gives $T_2 = 81 \pm 4 \,  \mathrm{\mu s}$ \cite{Carter:2015vc}. As $N_{\mathrm{V}}$ was not reported for this wafer, we assume the same creation yield as in Fig.~\ref{fig2}d and obtain $N_{\mathrm{V}} \sim 10^{15}  \, \mathrm{cm^{-3}}$. The longer coherence time is likely related to lower concentration of intrinsic paramagnetic centers and hence longer  $T_2^{\mathrm{(pristine)}}$ compared to that in our non-irradiated wafer (Fig.~\ref{fig5}b). The dashed line in Fig.~\ref{fig5}b  represents a fit to Eq.~(\ref{T2_Phi}) with $T_2^{\mathrm{(pristine)}} = 100 \, \mathrm{\mu s}$ and the same $\kappa$ as in the case above. 

Compared to electron irradiation, neutron irradiation creates a higher defect density, but also a higher percentage of  defects other than $\mathrm{V_{Si}}$, resulting in a larger $\kappa$. As a consequence, the coherence time is significantly shorter for a moderate concentration of the created $\mathrm{V_{Si}}$ centers. The coherence time measured for the neutron-irradiated samples should be seen as a lower bound, due to inhomogeneity in the applied microwave field across the detection volume. Application of optimized pulses will be required to test the limits of coherence in such samples \cite{Nobauer:2015kp}. Figure~\ref{fig5}b shows experimental data for HPSI 4H-SiC wafers irradiated in two scientific reactors with different distribution of neutron energy.  They can be described assuming $\kappa = 3.8 \times 10^{-11}  \, \mathrm{s^{ -1} cm^{3}}$ and $\kappa = 25.2 \times 10^{-11}  \, \mathrm{s^{ -1} cm^{3}}$  and the same $T_2^{\mathrm{(pristine)}} = 48 \, \mathrm{\mu s}$ as for our electron irradiated wafer, which is represented by the cyan area in Fig.~\ref{fig5}b. 

We now compare the aforementioned results to the spin coherence time in ion-irradiated samples. The ion irradiation can also damage the SiC lattice resulting in a reduction of $T_2$. It is expected that the irradiation induced damage is stronger for heavier atoms. From this point of view, it is advantageous to use light atoms --
 particularly H (protons) -- for the creation of highly coherent $\mathrm{V_{Si}}$ centers. In case of low proton irradiation fluence $\Phi_p = 2 \times 10^{11} \, \mathrm{cm^{-2}}$ corresponding to $N_{\mathrm{V}} = 3 \times 10^{13}  \, \mathrm{cm^{-3}}$ (Fig.~\ref{fig5}b), the coherence time $T_2 = 42 \pm 20 \, \mathrm{\mu s}$ approaches that in the pristine wafer $T_2^{\mathrm{(pristine)}} = 48 \, \mathrm{\mu s}$ \cite{Kraus:2017cka}.  It was found in another work that with increasing $\Phi_p$ and $N_{\mathrm{V}}$ the $T_2$ time decreases.  For $\Phi_p = 1 \times 10^{14} \, \mathrm{cm^{-2}}$ with estimated emitter density $N_{\mathrm{V}} = 3 \times 10^{14}  \, \mathrm{cm^{-3}}$, the spin coherence time $T_2 = 22 \pm 5 \, \mathrm{\mu s}$ from a stretched exponential fit was reported \cite{Embley:2017bf}. 

 \begin{figure}[t]
\includegraphics[width=.48\textwidth]{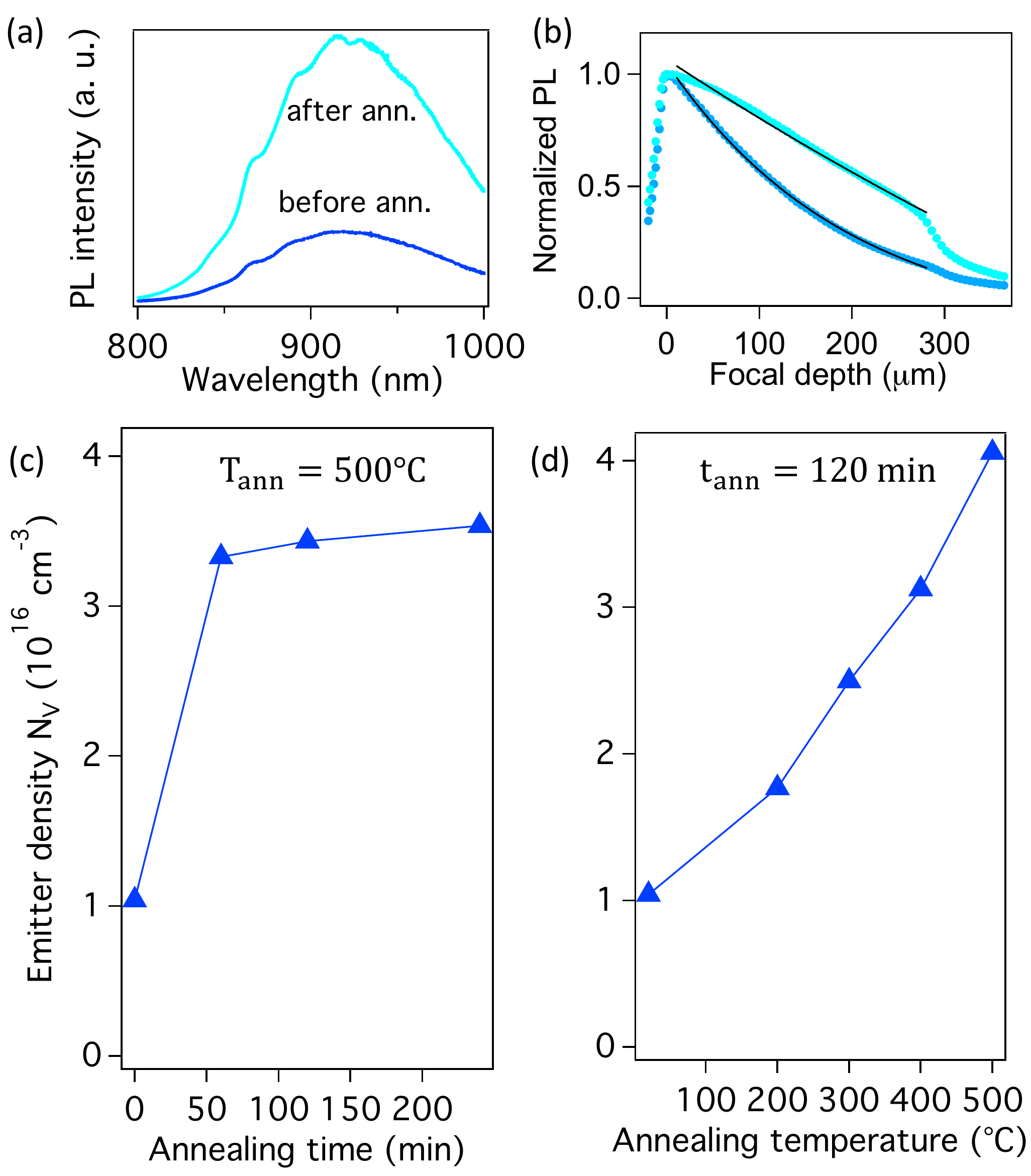}
\caption{Effect of annealing on the PL of a heavily neutron irradiated SiC crystal. The neutron irradiation fluence is $\Phi_n = 1 \times 10^{18} \, \mathrm{cm^{-2}}$. (a) Increasing of the PL intensity with annealing ($T_{\mathrm{ann}} = 500^{\circ}\mathrm{C}$ and $t_{\mathrm{ann}} = 240 \, \mathrm{min}$). (b) PL confocal scans before and after annealing with parameters as in (a). The solid lines are exponential fits. (c) Variation of the emitter density with  annealing time at $T_{\mathrm{ann}} = 500^{\circ}\mathrm{C}$. (d) Variation of the emitter density with annealing temperature for $t_{\mathrm{ann}} = 120 \, \mathrm{min}$. } \label{fig6}
\end{figure}
 
In fact, the dependence of the spin coherence on the proton (or any other ion) irradiation flucence is more complex as in case of electron or neutron irradiation. The ions penetrate to a certain depth, which depends on their energy and the ion type. Though the highest probability to create defects is at the Bragg peak,  there is a non-zero probability to create shallow $\mathrm{V_{Si}}$ centers, which have longer coherence time \cite{Brereton:2018ur}. We expect that Eq.~\ref{T2_Phi} can still be used for ion irradiation though the parameter $\kappa$ should depend on the ion type and energy as well as on the depth from the irradiated surface. 

\subsection*{Effect of annealing} 

Since, as previously mentioned, a high $\mathrm{V_{Si}}$ density combined with long spin coherence is desirable for various applications, we anneal a heavily irradiated sample to explore crystal healing. Figure~\ref{fig6}a shows the $\mathrm{V_{Si}}$ PL spectrum in a neutron irradiated sample with a fluence 
$\Phi_n = 1 \times 10^{18} \, \mathrm{cm^{-2}}$ and the corresponding initial emitter density $N_{\mathrm{V}} = 1 \times 10^{16}  \, \mathrm{cm^{-3}}$. Figures~\ref{fig6}c and \ref{fig6}d show the emitter density as a function of the annealing time $t_{\mathrm{ann}}$  and temperature $T_{\mathrm{ann}}$, respectively. The PL intensity linked to $N_{\mathrm{V}}$  increases by a factor of 4 after optimum annealing conditions. This increase can be explained by healing of some irradiation-induced and intrinsic defects other than $\mathrm{V_{Si}}$, which can provide non-radiative recombination paths for the $\mathrm{V_{Si}}$ defects if they are located in their vicinity. This conclusion is also confirmed by the in-depth (along the $z$ direction) PL scans, presented in Fig.~\ref{fig6}b.  Exponential fits to $\exp (- \alpha z)$ allow to estimate the absorption coefficient $\alpha = \alpha_{\mathrm{laser}} +  \alpha_{\mathrm{PL}}$. It decreases from $\alpha = 50.8  \, \mathrm{cm^{-1}}$ to $\alpha = 8.0  \, \mathrm{cm^{-1}}$ after annealing, indicating that some color centers are indeed removed. 
 
The annealing temperature in our experiments is limited to $500^{\circ}\mathrm{C}$ since the $\mathrm{V_{Si}}$ defects can otherwise be eliminated at higher temperatures \cite{Fuchs:2015ii}. We would like to note that we do neither observe any PL enhancement nor an inrease of $T_1$ or $T_2$ in the electron irradiated sample. This again strongly hints at less severe crystal damage and a higher percentage of created $\mathrm{V_{Si}}$ centers by electron irradiation.

\begin{figure}[t]
\includegraphics[width=.48\textwidth]{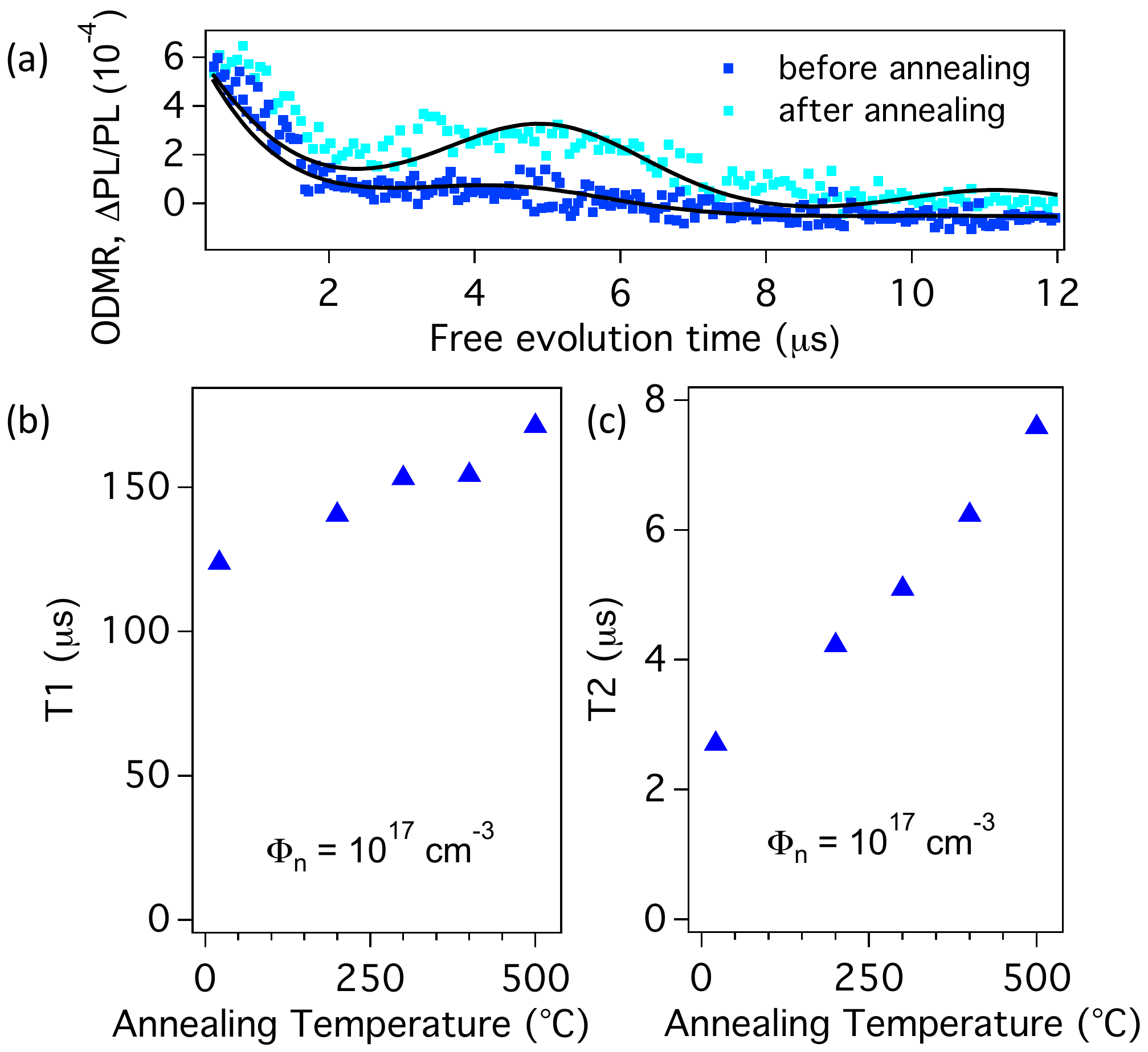}
\caption{Effect of annealing on the spin coherence properties of the $\mathrm{V_{Si}}$ defects. (a) ESEEM signal before and after annealing. (b) Evolution of the spin lattice relaxation time with annealing temperature. (c) Spin coherence time as a function of annealing temperature. The annealing time for each step in (b) and (c) is $t_{\mathrm{ann}} = 120 \, \mathrm{min}$. } \label{fig7}
\end{figure}

To explore the effect of annealing on the coherence properties of the $\mathrm{V_{Si}}$ defects, we measure $T_1$ and $T_2$ in the  $\Phi_n = 1 \times 10^{18} \, \mathrm{cm^{-2}}$ sample after each annealing step. Figure~\ref{fig7}a shows the spin-echo experimental data before and after annealing (symbols), fitted with ESEEM (the solid lines).  An improvement of the coherence properties with annealing can be clearly seen from these data. 

Figure~\ref{fig7}b presents the measured spin relaxation time $\mathrm{T_1}$ as a function of the annealing temperature ($t_{\mathrm{ann}} = 120 \, \mathrm{min}$). Though $\mathrm{T_1}$ increases, it is not completely restored to the 
value of $300 \,  \mathrm{\mu s}$ as in case of low irradiation fluences (Fig.~\ref{fig5}a). We note, $\mathrm{T_1}$ as a function of $T_{\mathrm{ann}}$ does not saturate in our experiments. For higher $T_{\mathrm{ann}}$, $\mathrm{T_1}$ might increase further, but $N_{\mathrm{V}}$ would drop due to healing out of the $\mathrm{V_{Si}}$ defects.  The spin coherence time $T_2$ can be improved by a factor of 2.5 as presented in Fig.~\ref{fig7}c. The maximum value achieved in the heavily irradiated sample is $T_2 = 7.6 \pm 0.5 \,  \mathrm{\mu s}$. Since $1 / T_2^{(i)}$ scales with the number of all created defects, this increase of $\mathrm{T_2}$ can be explained by healing of paramagnetic, non-$\mathrm{V_{Si}}$ defects at temperatures below $500^{\circ}\mathrm{C}$ and therefore a suppression of decoherence caused by the spin-spin interaction. 

To conclude, we have thoroughly investigated the irradiation impact on the room-temperature spin coherence properties of silicon vacancies in SiC.  We have measured the spin-lattice relaxation time and the spin coherence time depending on the irradiation particle (electron, neutron and proton), irradiation fluence and irradiation energy. In order to find the broader systematics, we have analyzed the available literature values, too.  

We have established, that the spin-lattice relaxation time remains constant up to high irradiation fluences, independently of the irradiation type. On the contrary, the spin coherence time is very sensitive to the irradiation type and fluence. The longest spin coherence time for the same emitter density is observed for electron irradiation. Surprisingly, the electron energy has no influence on the silicon vacancy coherence properties and all data can be well described by the same equation with two independent parameters. The shortest spin coherence time was observed in neutron-irradiated samples, which, however, can be partially recovered using thermal annealing. We thus expect that our study provides important information for the optimization of the $\mathrm{V_{Si}}$ defect spin coherence properties and allows to design quantum devices and structures with the desired parameters based on SiC as a platform.

\section*{Methods} 

Neutron irradiation was performed in two different research reactors. For the first irradiation set, we purchased a high-purity semi-insulating (HPSI) SiC wafer from CREE and used chambers DBVK and DBVR at the BER II reactor at Helmholtz-Zentrum Berlin. Since thermal neutrons produce negligible displacement, only epithermal and fast neutrons are counted in the fluence calculation \cite{Wendler:2012co}. The neutron fluence spanned a large range from $3.3 \times 10^{16} \mathrm{cm^{-2}}$ to $3.5 \times 10^{19} \, \mathrm{cm^{-2}}$. For the second irradiation set, we purchased a HPSI SiC wafer from Norstel and used the TRIGA Mark II reactor (General Atomics, San Diego) of the Atominstitut, TU Wien \cite{Fuchs:2015ii}. During irradiation, the temperature of the crystals did not exceed $100\,^{\circ}$C. Vacancy creation by neutron irradiation in SiC is weakly dependent on neutron energy in the range $0.18\,$MeV$ < E_n < 2.5\,$MeV, but falls off rapidly for smaller energies. The neutron energies in the TRIGA reactor are distributed evenly in the range between 100 eV and $1\times 10^5 \,$eV, with a peak in flux density around $2.5\,$MeV followed by a sharp cut-off.

For the electron irradiated sample series, a HPSI 4H-SiC wafer was
purchased from Norstel and diced in many pieces. The irradiation with electron energies of 1 and 2~MeV was performed  at 
electron beam irradiation facility at QST Takasaki (Japan). During the irradiation, the samples were placed on a water cooled copper plate to avoid heating by the electron beams. The irradiation with electron energies of 20 and 25~MeV was performed at the electron accelerator ELBE in HZDR (Germany). The beam was extracted through a 300 $\mathrm{\mu m}$ Beryllium vacuum window and passes 150 mm through air before hitting the sample. The incident beam distribution and the small angle scattering in the vacuum window and air leads to a Gaussian shaped beam distribution on the sample. The beam profile was measured at the position of the sample with a Roos ionisation chamber (IC) N34001. The certificate calibrated IC and a Faraday cup were used for the dose and current calibration. From the profile measurement, the irradiation parameters were deduced. To monitor the current stability of the electron gun as well as dark current contributions, the edge field of the beam was constantly measured and stabilized during the irradiation.

Proton irradiation was done at the TIARA irradiation facilities in Takasaki, Japan. A piece of a 4H-SiC wafer purchased from CREE was placed on an aluminum plate and irradiated with a focused proton beam generated by a single-ended particle accelerator with a typical spot size of $1 \, \mathrm{\mu m} \times 1  \, \mathrm{\mu m}$ and proton energy of $1.7\,$MeV. The irradiation beam current was monitored by a beam dump (Faraday cup) connected to a picoampermeter  before and after irradiation.

The continuous wave and pulsed ODMR measurements were performed with a confocal setup. For optical excitation, a 788 nm diode laser (LD785-SE400 from Thorlabs) was coupled into a 50 $\mathrm{\mu m}$ optical fiber and focused onto the sample using a $\times10$ objective (Olympus LMPLN10XIR), with the laser spot measuring approximately 10 $\mathrm{\mu}$m in diameter. The laser power at the surface of the sample was about 20 mW, illuminating a volume of approximately 300 $\mathrm{\mu}m^3$ for the measurements of electron and neutron irradiated samples. For the measurement of the proton irradiated sample, a $\times100$ objective (Olympus LCPLN100XIR) was used, which leads to laser illumination of approx. $2.6\ \mathrm{\mu}m^3$. The PL was collected through the same objective and separated from the scattered laser light using a 850 nm short pass dichroic mirror and a 875 nm long pass filter. Behind the filter, it was coupled into a 600 $\mathrm{\mu m}$ optical fibre and detected using a Si avalanche photodiode (APD120A from Thorlabs). The sample was placed on a 0.5 mm copper-stripline to apply the microwaves generated by a signal generator (Stanford Research Systems SG384) and amplifier (Vectawave VBA1000-18). A permanent magnet was mounted below the sample to generate the external magnetic field of 15 mT. To generate the laser pulses needed for time-resolved ODMR experiments, an acousto-optic modulator (Opto-Electronic MT250-A02-800) was placed in the laser beam path. The microwaves were modulated using an RF-switch (Mini-Circuits ZASWA-2-50DR+).  TTL pulsing for laser and microwave modulation was provided by a PulseBlaster ESR-PRO 500 MHz card from SpinCore. The ODMR  signal was processed using either a lock-in amplifier (Signal Recovery DSP 7230) or an oscilloscope card from GaGe CompuScope (Razor Max).

\section*{Acknowledgments}
This work has been supported by the German Research Foundation (DFG) under Grants  AS 310/5-1 and DY 18/13-1, the FWF project I 3167-N27 SiC-EiC as well as JSPS KAKENHI 17H01056 and 18H03770. 
HK states that part of this work was carried out at the Jet Propulsion Laboratory, California Institute of Technology, under a contract with the National Aeronautics and Space Administration; and acknowledges funding by the NASA Postdoctoral Program.
We thank M.~Villa and R.~Bergmann for operation of the TRIGA reactor as well as G.~Bukalis for operation of the BER II reactor.

\end{document}